\documentclass[10pt]{article}

\usepackage{amsmath}
\usepackage{amssymb}

\usepackage{graphicx}

\usepackage{cite}

\usepackage{color} 

\usepackage[labelfont=bf,labelsep=period,justification=raggedright]{caption}

\makeatletter
\renewcommand{\@biblabel}[1]{\quad#1.}
\makeatother

\date{}

\begin{document}

\begin{flushleft}
{\Large
\textbf{Individual rules for trail pattern formation in Argentine ants (\textit{Linepithema humile}).}
}
\\
Andrea Perna$^{1,2,3,4,\ast}$, 
Boris Granovskiy$^{2}$, 
Simon Garnier$^{3,4,5}$,
Stamatios Nicolis$^{2}$,
Marjorie Lab\'edan$^{3}$,
Guy Theraulaz$^{3,4}$,
Vincent Fourcassi\'e$^{3,4}$,
David J. T. Sumpter$^{2}$, 
\\
\bf{1} Complex Systems Institute of Paris Ile de France, 57-59 rue Lhomond, Paris, France 
\\
\bf{2} Collective Behaviour Group, Matematiska Institutionen, Uppsala Universitet, Box 480 - 75106 Uppsala, Sweden
\\
\bf{3} Centre de Recherches sur la Cognition Animale, UMR-CNRS 5169, Universit\'e Paul Sabatier, B\^at 4R3, 118 Route de Narbonne, 31062 Toulouse cedex 9, France.
\\
\bf{4} CNRS, Centre de Recherches sur la Cognition Animale, F-31062 Toulouse, France
\\
\bf{5} Department of Ecology and Evolutionary Biology, Guyot Hall, Princeton University - 08544 Princeton, NJ, USA
\\
$\ast$ E-mail: Corresponding author perna@math.uu.se
\end{flushleft}

\section*{Abstract}
We studied the formation of trail patterns by Argentine ants exploring an empty arena. Using a novel imaging and analysis technique we estimated pheromone concentrations at all spatial positions in the experimental arena and at different times. Then we derived the response function of individual ants to pheromone concentrations by looking at correlations between concentrations and changes in speed or direction of the ants.  Ants were found to turn in response to local pheromone concentrations, while their speed was largely unaffected by these concentrations. Ants did not integrate pheromone concentrations over time, with the concentration of pheromone in a 1 cm radius in front of the ant determining the turning angle. The response to pheromone was found to follow a Weber's Law, such that the difference between quantities of pheromone on the two sides of the ant divided by their sum determines the magnitude of the turning angle. This proportional response is in apparent contradiction with the well-established non-linear choice function used in the literature to model the results of binary bridge experiments in ant colonies (Deneubourg et al. 1990)\nocite{Deneubourg_et_al1990}. However, agent based simulations implementing the Weber's Law response function led to the formation of trails and reproduced results reported in the literature. We show analytically that a sigmoidal response, analogous to that in the classical Deneubourg model for collective decision making, can be derived from the individual Weber-type response to pheromone concentrations that we have established in our experiments when directional noise around the preferred direction of movement of the ants is assumed.

\section*{Author Summary}
Many ant species produce large dendritic networks of trails around their nest. These networks result from self-organized feedback mechanisms: ants leave small amounts of a chemical -a pheromone- as they move across space. In turn, they are attracted by this same pheromone so that eventually a trail is formed. In our study, we introduce a new image analysis technique to estimate the concentrations of pheromone directly on the trails. In this way, we can characterise the ingredients of the feedback loop that ultimately leads to the formation of trails. We show that the response to pheromone concentrations is linear: an ant will turn to the left with frequency proportional to the difference between the pheromone concentrations on its left and right sides. Such a linear individual response was rejected by previous literature, as it would be incompatible with the results of a large number of experiments: trails can only be reinforced if the ants have a disproportionally higher probability to select the trail with higher pheromone concentration. However, we show that the required non-linearity does not reside in the perceptual response of the ants, but in the noise associated with their movement. 

\section*{Introduction}
Many ant species produce large dendritic patterns of trails around their nests (fig.~\ref{fig:snapshots}). These patterns are among the most important examples of transportation networks built by animals: they mediate the exploration of space and the coordination of foraging activities across the whole colony~\cite{Hoelldobler_and_Wilson1990, Franks_et_al1991} and channel the daily movements of hundreds or thousands of ants. Empirical observations~\cite{Acosta_et_al1993, Buhl_et_al2009, Latty_et_al2011, Reid_et_al2011} and models~\cite{Sole_et_al2000, Berthouze_and_Lorenzi2008} have shown that ant trail networks provide efficient solutions for searching and transporting food.

In spite of having coherent and efficient organization on a large scale, ant trail formation can be explained as the result of a completely self-organized process. Simulations supported by experiments have shown that the trails are the result of an autocatalytic process: ants move in response to local concentrations of pheromone and in turn change these same concentrations by laying new pheromone where they go~\cite{Deneubourg_et_al1989, Calenbuhr_et_al1992, Calenbuhr_and_Deneubourg1992, Watmough_and_Edelstein-Keshet1995, Edelstein-Keshet_et_al1995, Helbing_et_al1997, Helbing_et_al1997a, Schweitzer_et_al1997, Couzin_and_Franks2003}. 

The link between models and experiment is weak in one important respect: the exact nature of how
individual ants move on the trail and respond to pheromone remains unknown. Important work in the direction of answering this question was done by Jean Louis Deneubourg and collaborators~\cite{Goss_et_al1989,Calenbuhr_and_Deneubourg1992,Beckers_et_al1992,Beckers_et_al1993,Jeanson_et_al2003},~\cite{Camazine_et_al2001}.
In Deneubourg's model, when individual ants face a bifurcating trail, their behaviour depends on pheromone concentrations on the trails ahead with a function of the form
\begin{equation}
P_L = \dfrac{(h+L)^a}{(h+L)^a + (h+R)^a} \; \textnormal{and} \; P_R = \dfrac{(h+R)^a}{(h+L)^a + (h+R)^a}
\label{eq:Deneubourg}
\end{equation}
where the probability $P_L$ ($P_R$) for an ant to select the left (right) branch of a bifurcating trail is expressed as a function of the concentrations of pheromone on the left~($L$) and right~($R$) branches. The parameter $a$ determines the degree of nonlinearity of the choice. A high value of $a$ means that even if one branch has only slightly more pheromone than the other, the ant will have a disproportionally large probability of choosing it. If $a = 1$ the ants react in a linear, proportional manner to pheromone concentration. The parameter $h$ acts as a threshold for response to pheromone. For larger values of $h$, more marking is necessary for the choice to become significantly non-random. 

Equation~\ref{eq:Deneubourg} (or variants of it) accounts well for the results of experiments in which ants face a branching point in their trail, usually in the form of a double bridge. For example, Deneubourg et al.~\cite{Deneubourg_et_al1990} studied a colony of Argentine ants (\textit{Linepithema humile}), which was given access to a bridge with two bifurcating branches leading to the same source of food. At the beginning of the experimental trials, ants took both branches in similar numbers, but ultimately selected one of the two bridges~\cite{Goss_et_al1989, Deneubourg_et_al1990}. Selection of one single branch is only possible if the choice function has a sigmoidal shape,i.e. when $a > 1$. The exact value of the exponent $a$ that reproduced experimental results the best changes for different experimental setups. In ref.~\cite{Deneubourg_et_al1990} it was found that an exponent $a \simeq 2$ reproduces experimental results, while another experiment with ants moving in a corridor found $a = 4$~\cite{Vittori_et_al2006}. Beckers et al.~\cite{Beckers_et_al1992, Beckers_et_al1993} went on to establish similar results for \textit{Lasius niger}, also finding an exponent $a \simeq 2$. This work has led to an established `wisdom' that ants react disproportionally (i.e. $a > 1$ ) to pheromone concentration. Equation \ref{eq:Deneubourg} with $a = 2$ has further been the basis of models of self-organised trail formation (e.g.~\cite{Deneubourg_et_al1989}) re-enforcing the idea that a disproportional response by ants at an individual level is required for trail formation.

Double bridge experiments do not however capture individual ant behaviour and are instead fitted directly to the global outcome. Many different mechanisms in which individual preferences are amplified by positive feedback can all explain the selection of one single branch (see e.g.~\cite{Stickland1993}) while little is known about the actual rules used by \emph{individual} ants to respond to pheromone concentrations. Questions such as how ants adjust their turning angle depending on pheromone concentrations, or whether they integrate pheromone information over a certain distance before committing to a decision, remain open.


How insects perceive and react to environmental stimuli has been studied in many contexts other than pheromone trails. For example, various studies have established that perceptual errors are proportional to the magnitude of the stimuli~\cite{Cheng_et_al1999, Sommer_and_Wehner2004, Shafir_et_al2005}. Such observations are what we expect if insect perception follows a Weber's Law~\cite{Weber1834}. A Weber's Law holds if the response $U$ to a difference between two stimuli $S_1$ and $S_2$ is proportional to ratios of the type ``signal difference''/``average signal''. In the case of stimuli for which $S_1$ and $S_2$ are equivalent and represented in similar proportion in the sensory field we have
\begin{equation}
U = c \frac{S_1 - S_2}{S_1 + S_2}
\label{eq:Weber}
\end{equation}
where $c$ is a constant. The ratio $\left( S_1 - S_2 \right)/\left( S_1 + S_2 \right)$ in equation~\ref{eq:Weber} is known as the ``Michelson contrast'' of the stimulus pattern. Weber's Law is known to hold for insects (e.g. \cite{Smith_and_Getz1994}), and is particularly well studied for human perception in different sensorial modalities (musical pitch, sound loudness, image brightness, length, speed, shape, time and numerosity). Equation~\ref{eq:Weber} is a linear function of difference and is thus analogous to a version of equation \ref{eq:Deneubourg} in which $a = 1$ (see methods). 

Studying individual responses to the trail pheromone is non-trivial since pheromone density is not readily visible. Two possible approaches are to test the ants' response to synthetic pheromone \cite{VanVorhisKey_and_Baker1982, VanVorhisKey_and_Baker1982a, Suckling_et_al2008} or to place strips removed from trails formed under controlled conditions \cite{Robinson_et_al2005, Jackson_et_al2006}. The former of these approaches allows us to control concentration, but it does not reproduce the physical and physico-chemical properties of the trails. It is difficult to relate the concentrations of synthetic pheromones to the actual concentrations present on the trail. For instance, pure (Z)-9-hexadecenal -the active compound of Argentine ant trail pheromone- is 200 times less active when presented alone than in the form of ant gaster extracts with an equivalent amount of molecule \cite{VanVorhisKey_and_Baker1982}. 
Trails formed on strips capture physical characteristics of the trail but do not allow us to tightly control concentrations. In addition, trails on paper strips are necessarily one-dimensional and cannot be easily used to study pheromone response in two dimensions. 

In our study we use novel imaging and analysis techniques to overcome these limitations and study trail following behaviour over a large number of tracking events. We assume that the concentration of pheromone at a particular position in the arena at a particular time is proportional to the number of ants that have previously passed that position. We use various assumptions about evaporation to test the robustness of our results. In this way, we can derive the response function of individual ants to pheromone concentrations directly on the trails produced by the ants themselves.

\section*{Results}

\subsection*{Arena-level observations of trail formation}
\begin{figure}[!ht]
\begin{center}
\includegraphics[width=0.4\textwidth]{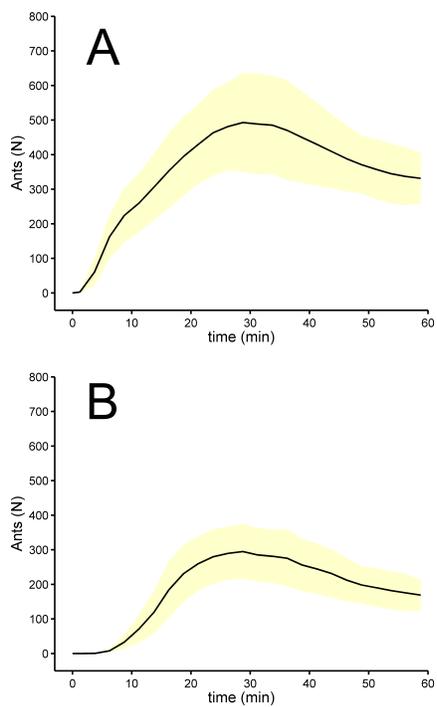}
\end{center}
\caption{
{\bf Arena level statistics of exploration.}  A. Number of ants in the arena over time. B. Number of ants along the arena border (i.e. less than 2.5 cm from the border) over time. For each plot the curve gives the mean and standard deviation over all trials.
}
\label{fig:arena_level_stats}
\end{figure}

The ants explore the arena uniformly in all directions around the entrance in the beginning of the experiment, but soon start to form trails that persist for some time. Later in the experiment these trails are either abandoned or amplified (see figure~\ref{fig:snapshots}). The kinetics of arena exploration is shown in figure~\ref{fig:arena_level_stats}. The figure reports the total number of ants in the arena (figure~\ref{fig:arena_level_stats}A) and of those along the arena border (figure~\ref{fig:arena_level_stats}B) as a function of time from the beginning of the trial. Ants start entering the arena soon after they are given access to it and start concentrating along the border of the arena about 10 minutes later. Usually the number of ants in the arena reaches a maximum about 30 minutes after the beginning of the experiment and then decreases slightly during the rest of the experiment. Roughly at the same time as the number of ants in the arena reaches its maximum, the number of trails also attains a maximum. 

\subsection*{Individual-level behaviour}
\begin{figure}[!ht]
\begin{center}
\includegraphics[width=0.8\textwidth]{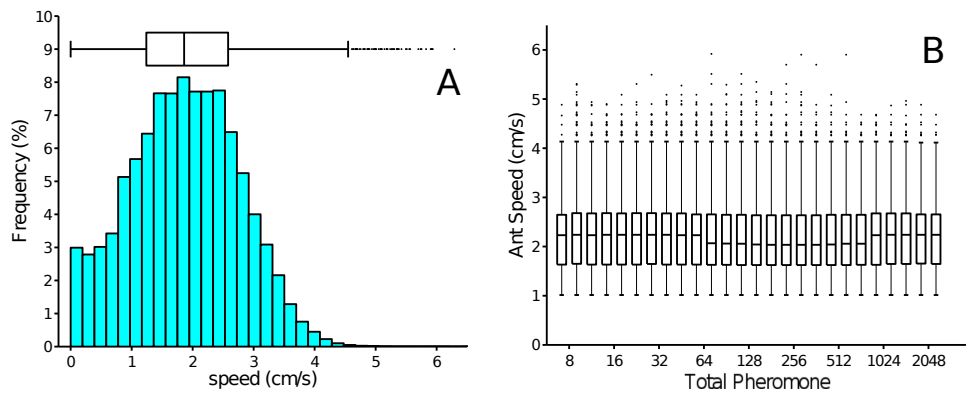}
\end{center}
\caption{
{\bf Speed distribution for individual ants (plot A) and distribution of speed as a function of the total pheromone encountered $(L+R)$ (plot B)}. The graph in A is from all data, while the one in B is limited to ''moving`` ants (ants that move at least 0.4 cm in each of the two time intervals; see methods) because of the difficulty in defining regions L and R for ants that do not move. The speed measured in both graphs is the average speed over 0.4 second intervals. The median and other percentiles are affected by small quantization effects, because ant position is recorded in pixel coordinates (1 pixel $\sim$ 0.8 mm, with small differences in different trials). The sample sizes of the boxplots in B are different. The whiskers in all the boxes represent data within 1.5 IQR of the quartiles; circles represent outlier data points.
}
\label{fig:speed}
\end{figure}

A first characterisation of individual ant motion is provided by the measure of ant speed (figure~\ref{fig:speed}A). Speed, measured over time intervals of 0.4 seconds for all the 600,000 tracking events, averaged to slightly less than 2~cm/s. The distribution of speeds is wide, and a few ants were observed to move at speeds of up to~6 cm/s. 

\begin{figure}[!ht]
\begin{center}
\includegraphics[width=0.3\textwidth]{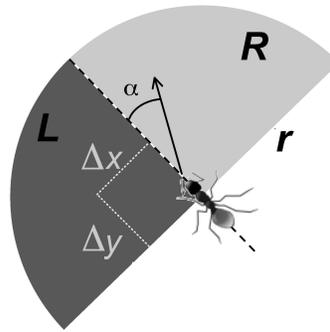}
\end{center}
\caption{
{\bf Regions used for estimating the concentrations of pheromone around the ant.} For each tracking event we get the position of the ant at time $t$ and its direction at times immediately prior to or after $t$ (average directions during 0.4 s intervals). The angle $\alpha$ is the change from previous direction. $L$ and $R$ are the integrals of all pheromone in two circular sectors ahead of the ant on the left and right side, respectively.
In order to avoid spurious correlations between ant movement and the pheromone added by the ant during that same movement, we always calculate correlations with the pheromone map as it was 16 seconds before the tracking event (tmap = t – 16).
}
\label{fig:regions_R_and_L}
\end{figure}

We tested whether speed depends on pheromone concentrations experienced by the ants. Figure \ref{fig:speed}b gives box plots of ant speeds measured over the 0.4 seconds immediately after the ant experienced a given pheromone concentration in the two sectors of radius 1 cm $L$ and $R$ as illustrated in figure \ref{fig:regions_R_and_L} (see methods). The speed is clearly not influenced by pheromone concentration at this short time scale. This does not exclude the possibility that –over a longer time scale ants move faster on marked than on unmarked substrates. For example, a higher turning rate on substrates with little pheromone might result in lower distance travelled than when walking directly down a well-marked trail.

\begin{figure}[!ht]
\begin{center}
\includegraphics[width=1\textwidth]{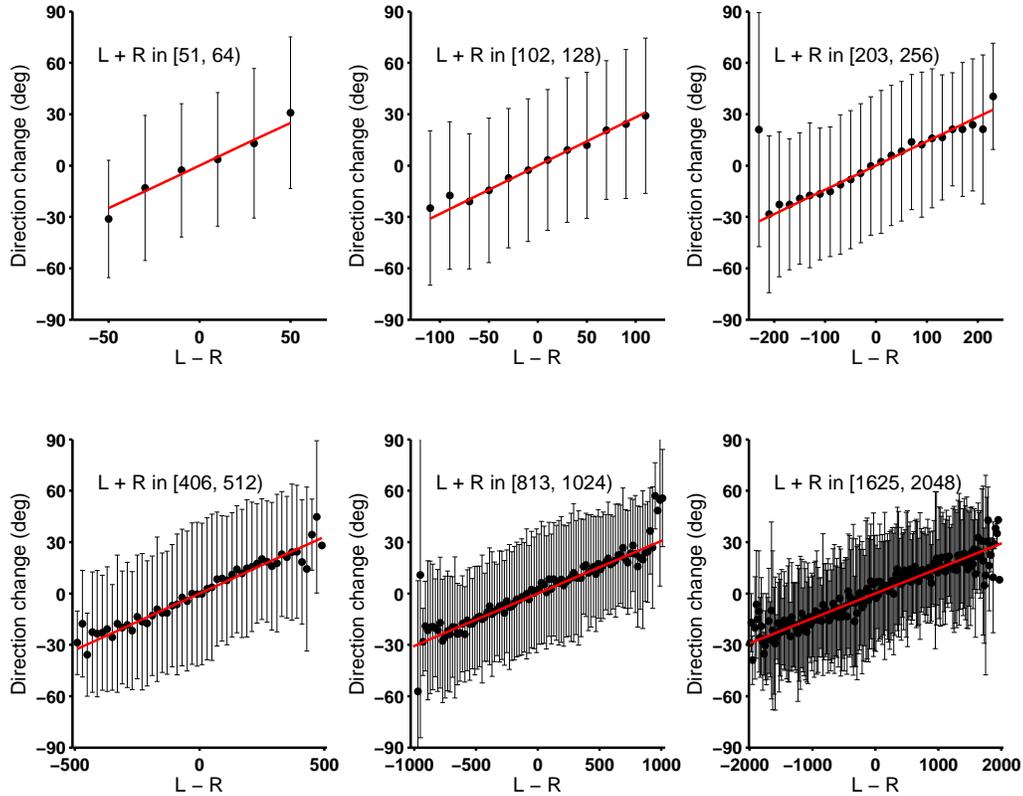}
\end{center}
\caption{
{\bf Measured angles of changes in direction made by ants as a function of pheromone difference $(L - R)$.} Each graph is for a different range of values of total pheromone $(L + R)$. Given the large number of data points involved in the plot, and to improve visualization, we only report the mean and standard deviation of data binned in intervals of 20 pheromone units. The red line is a linear fit (on the unbinned data) of the form $y=kx$. Statistics on the fitted slope $k$ are as follows: (A) $k = 0.4981$; $F_{33953,1} = 741$ , $p<0.001$; $R^2 = 0.0214$. (B) $k = 0.2820$; $F_{27112,1} = 1171$, $p<0.001$; $R^2 = 0.0414$. (C) $k=0.1416$; $F_{22152,1} = 1836$, $p < 0.001$; $R^2 = 0.0766$. (D) $k = 0.0662$; $F_{17799,1} = 2303$, $p < 0.001$; $R^2=0.1146$. (E) $k = 0.0307$; $F_{17591,1} = 3558$, $p < 0.001$; $R^2 = 0.1682$. (F) $k = 0.0146$; $F_{10864,1} = 2102$, $p < 0.001$; $R^2 = 0.1621$.  Angles increase anticlockwise: positive angles indicate an ant turn to the left. The data from all the trials are merged for this figure.
}
\label{fig:alpha_vs_L_minus_R}
\end{figure}

\begin{figure}[!ht]
\begin{center}
\includegraphics[width=1\textwidth]{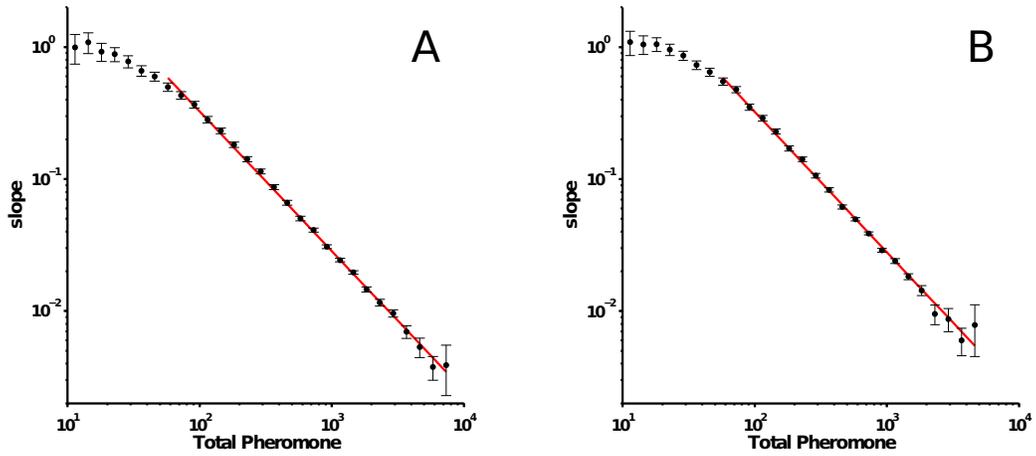}
\end{center}
\caption{
{\bf Log-log plot of the slope $k$ of the angle change $k = \alpha / (L - R)$ vs. the total pheromone around one ant $(L + R)$. } Error bars associated with each data point are $95\%$ confidence intervals on the slope estimation.  A. no pheromone evaporation. B. pheromone evaporation with half-life of 30 min. The red line is a power-law fit of the form   (non-weighted linear least squares of the log-transformed data; fitting parameters for the curve in A: $A = 42.41 (95\% CI: 36.90, 48.74); \beta = 1.058 (95\% CI: 1.079, 1.037); R^2 = 0.9982; F_{20,1} = 11084; p < 0.001$. curve in B: $A = 43.46 (95\% CI: 34.13, 55.34); \beta = 1.064 (95\% CI: 1.102, 1.026); R2 = 0.9949; F18,1 = 3489; p < 0.001$); fit restricted to the data point with $L+R > 50$ pheromone units. 
Assuming $\beta = 1$ and fitting directly a function of the form $\alpha = A_0 \left(L-R\right)/\left(L+R+T_0\right)$ to all the original data, where $T_0$ is a threshold for pheromone detection, gives the best fitting values $A_0=30.80$ and $T_0 = 10.53$ for the condition without evaporation and $A_0=30.61$ and $T_0 = 8.61$ for the condition with evaporation.
}
\label{fig:slope_vs_L_plus_R}
\end{figure}

Let us consider the angle changes made by ants in response to pheromone concentrations. For each level of total pheromone $(L + R)$ we look at how the average turning angle depends on the difference between the two concentrations of pheromone on the left and on the right, i.e. $(L - R)$. Figure \ref{fig:alpha_vs_L_minus_R} shows this plot for six different values of total pheromone. Here (and in figure~\ref{fig:slope_vs_L_plus_R}) individual tracking data from all the trials are combined together (table~\ref{tab:no_evaporation} provides statistics for the individual trials when assuming no significant pheromone evaporation and table~\ref{tab:with_evaporation} reports the same data when pheromone evaporation is assumed). Ants tend to turn in the direction of higher pheromone concentration. For all concentrations of total pheromone $(L+R)$, the turning angle increases linearly with the difference between pheromone on the left and on the right of the ant
\begin{equation}
\alpha = k \left( L - R \right)
\label{eq:alpha_linear}
\end{equation}
The slope $k$ of the linear relation changes from one plot to the other. In particular, the maximum turning angle is always of the order of $\pm35$ degrees, occurring when all the pheromone is concentrated on one side of the ant. This also coincides approximately with the typical standard deviation in the angle given the fit of equation~\ref{eq:alpha_linear}, indicating that 35 degrees is the relevant scale for changes in direction for our chosen observation scale.

In order to visualize how the slope $k$ changes with the total pheromone concentration $(L+R)$, figure~\ref{fig:slope_vs_L_plus_R} plots $k$ versus $\left( L + R \right)$. For a wide range of pheromone concentrations, the values of $k$ follow a straight line in the log-log plot. This indicates a power-law relationship of the type
\begin{equation}
\label{eq:power_law}
k = A \left( L + R \right)^{-\beta}
\end{equation}
for some constant $\beta$, or substituting equation~\ref{eq:alpha_linear}
\begin{equation}
\label{eq:alpha_complete}
\alpha = A \left(L-R \right) \left(L+R \right)^{-\beta}
\end{equation}
The curve bends down for low concentrations of total pheromone ($(L + R) < 50$ pheromone units). This probably reflects the fact that ants cannot sense pheromone concentrations below a certain threshold. This can happen because the substrate was not properly marked (because real ants mark intermittently~\cite{Wilson_and_Pavan1959} and could have skipped that patch) and/or because the concentration of pheromone is lower than the minimal amount that one ant can sense (the sensory detection threshold of the ant). If the ant does not sense the pheromone, we expect it on average to continue moving forward without changing direction, i.e. we expect the curves of figure~\ref{fig:slope_vs_L_plus_R} to go to zero when $L + R$ is equal to the detection threshold. Assuming a threshold of 50 pheromone units, we can fit equation \ref{eq:power_law} directly to all 12 trials (table \ref{tab:no_evaporation}). In all cases $\beta \simeq 1$ and when all trials are combined the best-fit slope is $\beta = 1.06$. For $\beta = 1$, equation \ref{eq:power_law} becomes
\begin{equation}
\label{eq:alpha_Michelson}
\alpha = A \dfrac{\left(L-R \right)}{ \left(L+R \right)}
\end{equation}
which expresses a relation between pheromone concentrations and turning angles of the same type we would expect based on Weber's Law.

It is important to notice that this relation does not depend on the specific scale used to measure pheromone: the same relation holds if we multiply both $L$ and $R$ by the same constant. This makes the result independent of the assumption used in the methods that ants mark one square millimetre of arena with one unit of pheromone every one second: whatever the amount of pheromone laid down per time step, we will find the same result. Another implication of the form of equation~\ref{eq:alpha_Michelson} is that this result is not likely to be strongly affected by pheromone evaporation. Evaporation decreases pheromone in proportion to its quantity, on average producing an effect similar to multiplying both $L$ and $R$ by some constant smaller than one. In fact, assuming an evaporation rate of pheromone with $\lambda = 30 $min leads to a curve (in figure~\ref{fig:slope_vs_L_plus_R}B) very similar to the one obtained without evaporation at all (in figure~\ref{fig:slope_vs_L_plus_R}A). Nonetheless, evaporation at a much faster rate would probably have an effect because pheromone quantity drops below the detection threshold.

We also made the assumption that pheromone marking is reasonably constant in time. This implies that the values of $L$ and $R$ on which we base the analysis are proportional to the actual pheromone concentrations. We know that the marking rate of real ants is not constant, but it is likely to vary both because of random fluctuations and inter-individual variability. However, provided this variability is uncorrelated with pheromone concentrations it will only add noise to the data, but won't change our results. It remains to be thoroughly tested whether ants might modulate their trail laying behaviour depending on the concentration of already present pheromone. The existing evidence does not however support such a hypothesis. Aron and collaborators~\cite{Aron_et_al1993} report that Argentine ant colonies keep marking the substrate over time, but they do not test if the marking frequency (or the amount of pheromone per marking event, which is more difficult to test) remains constant. In his master thesis, Gerbier describes a control experiment for the paper~\cite{Gerbier_et_al2008}. In this experiment, Argentine ants were allowed to move between a nest and a food source through a narrow bridge. Focusing on two cm of the bridge, he measured the proportion of ants marking with pheromone, both 10 and 20 minutes after the beginning of the experiment. For ants directed towards the food source (which can be better compared to our experiments, in which there is no food at all in the arena) he measured a proportion of marking ants of $0.31 \pm 0.03$ after 10 minutes and $0.26 \pm 0.03$ after 20 minutes: in spite of the fact that the ant traffic in these conditions is much higher than in our experiments (because the whole colony is foraging for food along a narrow bridge) the proportion of marking ants did not change significantly over time.  Earlier experiments by Beckers and collaborators~\cite{Beckers_et_al1992} on $Lasius niger$ had also found ``that negative feedback between the trail strength and the trail laying is not so clearly in evidence''. Even if these studies are not directly comparable to ours because they involve the presence of food, it is reasonable to believe that if ants were to modulate their trail laying behaviour, they would be more likely to do it when there is food than when there is no food in the arena, and for this reason we should not expect such a modulation of pheromone deposition in our experimental conditions.

For the sake of our analyses, we assumed that ants respond to pheromone in front of them and up to a distance of one centimetre. We still have no real knowledge of what regions around their position ants actually use to detect pheromone. This raises several questions. For instance, we would like to know the perception radius over which ants can respond to pheromone, or at which points in front of the ant pheromone concentrations have the greatest effect on movement decisions. We would also like to check whether locations behind the ant are important. If this were the case, it would imply integration of pheromone concentrations over some time before deciding to change direction. 
%

\begin{figure}[!ht]
\begin{center}
\includegraphics[width=0.5\textwidth]{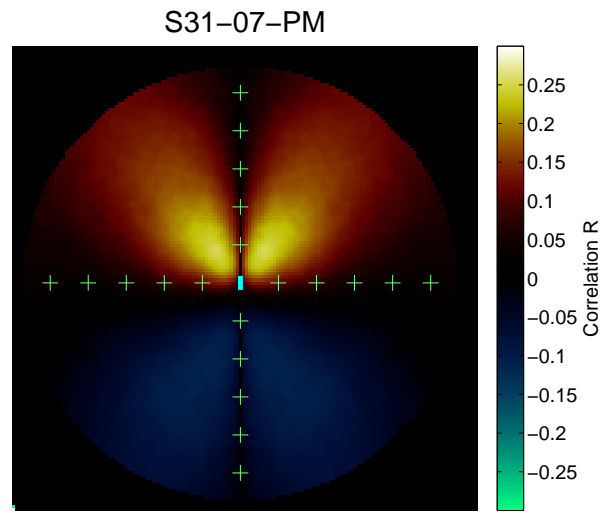}
\end{center}
\caption{
{\bf Correlation map between the observed turning angle $\alpha$ and the angle $\langle \alpha \rangle$ predicted using pheromone information at positions $\Delta x$ and $\Delta y$ (equation \ref{eq:alpha_Michelson_point_to_point}).}  The colour associated with each point ($\Delta x$, $\Delta y$) represents the value of the correlation coefficient between the observed turning angle $[\alpha_1, \alpha_2, …, \alpha_N]$ and the turning angles predicted from equation \ref{eq:alpha_complete} $[\langle \alpha_1 \rangle  , \langle \alpha_2 \rangle , ... , \langle \alpha_N \rangle  ]$. The ant is situated in the centre of the map, facing upwards, and its approximate dimensions are given by the cyan rectangle. The scale for the figure is provided by + symbols, which are spaced 1 cm. apart. The map is for trial T09 and no pheromone evaporation. Similar maps are found in all the trials.
}
\label{fig:correlation_map}
\end{figure}

In order to explore these issues we recomputed the predicted turning angle, but this time instead of integrating over the entire regions $R$ and $L$ we calculated $L_{\Delta x, \Delta y}$ and $R_{\Delta x, \Delta y}$ for each combination of two pixels at positions $\Delta x$ and $\Delta y$ relative to the position of the ant (see figure~\ref{fig:regions_R_and_L}). $\Delta x$ is always positive, while $\Delta y$ is either positive or negative, depending on whether we are looking at points in front or behind the ant. We then calculated
\begin{equation}
\label{eq:alpha_Michelson_point_to_point}
\langle \alpha \rangle = A \dfrac{L_{\Delta x, \Delta y} - R_{\Delta x, \Delta y}} {L_{\Delta x, \Delta y} + R_{\Delta x, \Delta y}}
\end{equation}
for each distance combination $\Delta x$ and $\Delta y$. For each pixel pair we measured the correlation coefficient $r_{\alpha,\langle \alpha \rangle}$ between the real turning angles $\alpha$ observed in all tracking events and the angles $\langle \alpha \rangle$ predicted by equation \ref{eq:alpha_Michelson_point_to_point}. Figure~\ref{fig:correlation_map} reports the resulting correlation map for one replicate of the experiment. The maximum correlation is found at about one cm from the ant position and around $\pm 45$ degrees from the heading direction. This value could be compatible with the position of the antennae. Care should however be taken in interpreting this result, since for small $\Delta x$ the pheromone field is necessarily highly correlated. In other words, $L_{\Delta x, \Delta y} - R_{\Delta x, \Delta y}$ is small in such cases. 

Negative correlations behind the ant are small, and can probably result from sensory adaptation at the level of the antennae: the left and right antenna would underestimate the amount of pheromone if they have just been exposed to high pheromone concentrations. The correlation becomes positive around the position of the ant, indicating a small temporal delay between perception and response (absence of integration). Given the measured ant speed of about 2~cm/s, if ants required a processing time of, for example, 0.5 seconds from the moment when they sense the pheromone until the moment when they change direction, then the correlation should become positive already $\sim 1$ cm behind the position of the ant, which is not observed.

\subsection*{Modelling collective patterns}
\begin{figure}[!ht]
\begin{center}
\includegraphics[width=1\textwidth]{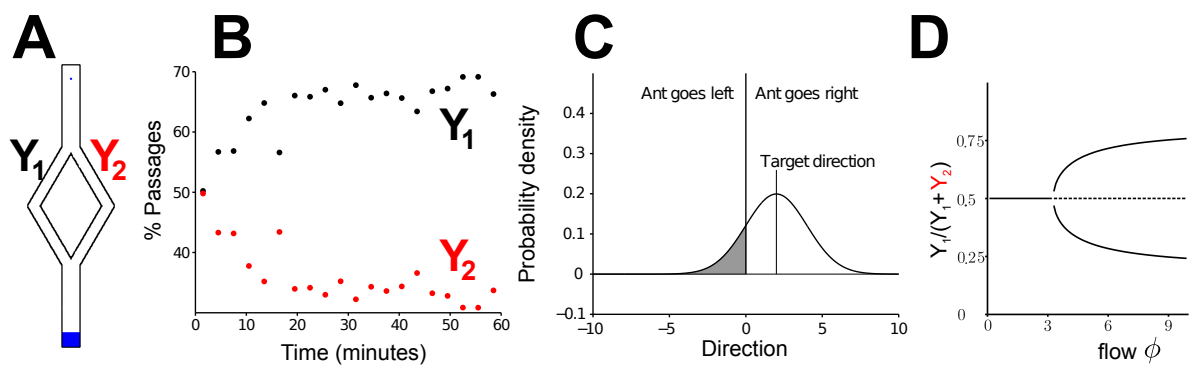}
\end{center}
\caption{
{\bf Simulation and analytical results implementing Weber's Law type response to pheromone in a binary bridge.}   \textbf{A}. Illustrative drawing of the simulation domain; the blue dot near the top of maze represents the nest, while the large blue region at the bottom is the “food source”. \textbf{B}. Percentages of simulated ants on each branch of the maze at different times in one run of simulation (each point represent the average over three minutes of simulation). C. Schema providing an intuitive explanation of equation \ref{eq:erf}. The target direction of one ant depends linearly on $(L – R)$. The probability for the ant to choose the left branch depends on the target direction and the directional noise. More precisely, if we assume that the branching point between left and right branch is at direction zero, the probability that the ant chooses the left branch is given by the integral of the curve in figure B from $-\infty$ to 0. D. Bifurcation diagram for the density of ants on one branch of the bridge ($\textnormal{Y}_1$) as a function of the total flow of ants in the setup.
}
\label{fig:results_double_bridge}
\end{figure}

The response rules of individual ants to pheromone concentrations established here are, at first sight, different from the disproportional response (equation~\ref{eq:Deneubourg}) found in the literature for large-scale binary choice experiments. To explain this discrepancy we ran an agent based simulation in a binary bridge setup where each agent (each ant) responds to pheromone according to the same rules identified from the experiments  (see methods). Figure~\ref{fig:results_double_bridge}B reports the percentage of ants on each of the two branches for one run of the simulations. The selection reaches a plateau at about 70\% of the ants taking one of the two branches. This response can be made stronger or weaker depending on the size of the bridge, the noise term and the interaction zone of the ant. A quantitative comparison with experiments reported in the literature is not directly possible, as our parameters for ant movement were estimated in an open arena, where thigmotactic responses are not important. These responses are no longer negligible in a binary bridge setup. However, the simulations indicate that a Weber's Law response to pheromone concentrations is compatible with the collective selection of one single branch of the bridge. 

To see why a proportional response inherent in Weber's law produces a disproportional outcome, consider the case of one ant approaching the branching point in which there is slightly more pheromone on the right than on the left and the total pheromone concentration is low (i.e. $R > L$ and $L + R$ small). Assuming that the journey towards the branching point is a random walk with a bias proportional to equation~\ref{eq:alpha_Michelson} (toward the ``target direction'' in figure~\ref{fig:results_double_bridge}C), then the point at which the ant reaches the branching point will be determined by a Normal distribution around the target direction. The probability of arriving at a point left of the branching point is then given by the integral of the Normal distribution over all values on the left of the branching point (from $-\infty$ to $0$ in figure~\ref{fig:results_double_bridge}C):
\begin{equation}
\label{eq:erf}
P_{L}(t) = \frac{1}{2} \left[ 1+ \operatorname{erf}\left( \frac{(L-R) t}{(L+R) \sqrt{4 D_2 t}}\right) \right]
\end{equation}
where $t$ is the number of steps in the random walk before the branching point and $D_2$ is a measure of the directional error in the ant movement. A more formal derivation of equation~\ref{eq:erf} is provided in the methods. The error function (erf) is, like equation~\ref{eq:Deneubourg}, a sharply increasing non-linear function. For practical purposes, both equation~\ref{eq:erf} and equation~\ref{eq:Deneubourg} would provide equally good fits to double bridge data and have similar mathematical properties. Figure~\ref{fig:results_double_bridge}D plots the bifurcation diagram for the density of ants on one of the two branches of a binary bridge when the individual choice function is of the form of equation~\ref{eq:erf} (see methods). This diagram is similar to the one obtained from equation~\ref{eq:Deneubourg}.

\begin{figure}[!ht]
\begin{center}
\includegraphics[width=1\textwidth]{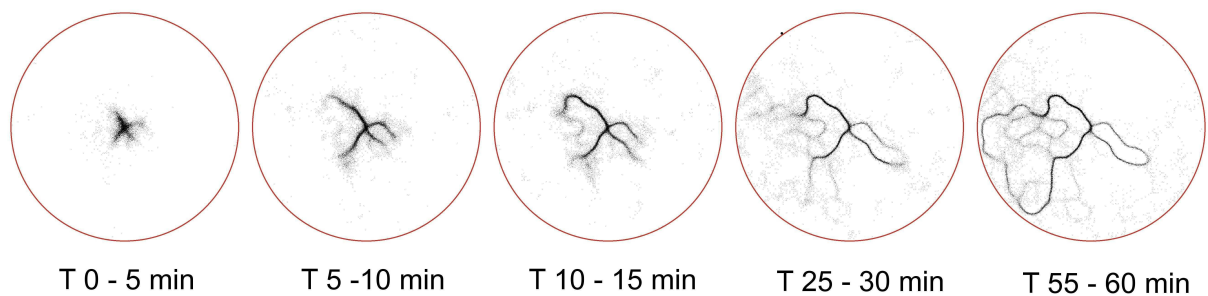}
\end{center}
\caption{
{\bf Output of the multi-agent simulation implementing the individual level parameters observed experimentally.}  Each image is obtained by summing 300 snapshots of the simulation taken at equal intervals of 1 second of simulation time (corresponding to 5 minutes of simulation) in a similar way to what had been done for the experimental data.
}
\label{fig:simulation_open_arena}
\end{figure}

When the simulation is run with the same parameters in a circular open arena, it leads to the formation of distinct trails (figure~\ref{fig:simulation_open_arena}). The timescale for the formation of trails is similar in the simulation and in the experiments, although the trails start appearing a bit earlier in the simulations. This discrepancy is in part because real ants show some latency for entering the arena at the beginning of the experiment. The most important difference between real and simulated trails is that simulated paths are more winding and more prone to form loops than real ones. Loops can form in the experiments as well. They are often stable for a long time, but usually they result from two trails connecting the nest to the arena border and a third one forming an arc of a circle around the border of the arena, a region where the density of ants is always high (see figure~\ref{fig:snapshots} and figure~\ref{fig:arena_level_stats}-B).

\section*{Discussion}
Weber's Law has previously been established in a wide range of animals and for different sensory stimuli~\cite{Stevens1957, Johnson_et_al2002}. Using such an internal metric allows animals to process different sensorial stimuli over a wide range of scales using a limited number of neurons~\cite{VanHateren1992}. The presence of Weber's Law for Argentine ants is partially explained by their need to respond to pheromone at very different concentrations~\cite{Pasteels_et_al1986}. Weber's Law implements a mechanism of perception where the response only depends on the ratio between pheromone concentrations $L$ and $R$: multiplying both $L$ and $R$ by the same number does not change the ants' response. Pheromone evaporation essentially corresponds to multiplying both $L$ and $R$ by the same constant smaller than one. Hence, Weber's Law-like mechanisms offer the additional advantage of determining stable responses even in the presence of evaporation. 

We have shown that the use of Weber's Law also explains the formation of ant trail networks. In the context of social interactions, the perceptual response is coupled with positive feedback to generate collective patterns. In our case, positive feedback is mediated through leaving pheromone, and the collective pattern is the trail network. We can imagine that other collective phenomena, such as group decision-making, could also be founded on coupling between Weber's Law and simple feedback mechanisms.

In contexts where groups of animals are faced with choosing between multiple options, such as the shortest path to food or the best direction to move~\cite{Sumpter_et_al2008, Ame_et_al2006}, disproportional non-linear responses are often implicated~\cite{Sumpter_and_Pratt2009, Deneubourg_et_al1990, Dussutour_et_al2005}. The Weber's Law we have established here is essentially a proportional linear response to stimulus differences. Indeed it is similar to a Deneubourg model (equation~\ref{eq:Deneubourg}) with $a = 1$. This raised a potential contradiction between our current results and earlier double bridge and binary choice experiments, which have established $a > 1$. We can explain this apparent contradiction by noticing that the actual end point of one ant approaching a branching point depends not only on the pheromone concentrations, but also on the directional noise. The probability of entering the left or right branch of a binary bridge is then obtained by integrating over all the possible outcomes on the left and on the right sides of the bifurcation. This integral is a non-linear error function, which is in many respects equivalent to a Deneubourg model for collective decision.

\begin{figure}[!ht]
\begin{center}
\includegraphics[width=1\textwidth]{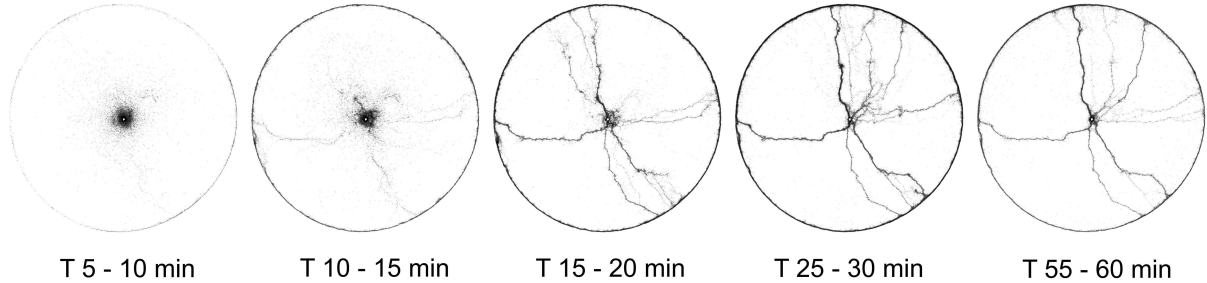}
\end{center}
\caption{
{\bf Evolution of the pattern formed by one colony (T09) over time.} Each picture is obtained by summing all the ants detected from arena-level snapshots during 5 minutes (300 snapshots). The contrast and gamma are adjusted to make single ants visible in the images.}
\label{fig:snapshots}
\end{figure}

While our observations are consistent with double bridge experiments, we do not have a complete explanation for the formation of trail networks. Comparing figures~\ref{fig:snapshots} and~\ref{fig:simulation_open_arena} we see several differences. The general structures of the trail networks are similar, but the trails obtained in the simulations appear earlier, are more winding and as a result contain more loops. By tuning the error we can improve some of these properties, but usually at the expense of realism in some other property. More specifically, the formation of trails in the experiments is often preceded by a short phase of isotropic exploration around the entrance (see figure \ref{fig:snapshots}). In the simulations, this pattern can be reproduced by increasing the error. However, doing so makes the trails appear later and can produce extremely winding trails that form loops just around the entrance. Our model shows to what degree trail networks can be explained by the ant response to a single pheromone, leading us to conclude that this response alone is not sufficient to completely explain the structure of the trails. Even if \textit{Linepithema humile} ants are known to rely predominantly on chemical signals for orientation~\cite{Aron_et_al1993}, we can imagine that they complement information carried by pheromone with more global inferences of their position and direction of movement. Many mechanisms such as path integration or klinotaxis can provide this additional information. Path integration in particular could explain some of the differences between real and simulated trails. In fact, the simulations are a direct implementation of the rules of response to pheromone measured experimentally over single time steps. As such, they reproduce exactly the pattern of movement of ants at the time scale of one time step. At a longer time scale, however, the trails produced in simulation bend in different directions, whereas the real trails are straighter. Path integration would imply that the turning angles of the ants are anti-correlated at longer time scales, ensuring that they remain more or less straight in spite of the noisier movement of ants at single time steps. This would effectively produce trails that preserve the same local structure, but are straighter at a larger scale. For a subset of the data we had individual ant traces over relatively longer time. From these traces we looked at the correlation between the turning angle of one ant at time $t$ and the turning angle at time $t +\Delta t$. This analysis showed a slight positive correlation over a short time scale ($\Delta t$ small): an ant which is turning to the left now will quite likely be turning to the left also one or two seconds later. These correlations decayed for longer time intervals, but never inverted of sign, indicating that there is no ``oscillatory'' movement around a trail with a fixed periodicity, though the ants might still have an increased probability of turning in the opposite direction after a variable interval of time. However, the set of data for which reliable individual tracking was available over multiple time steps was relatively small and did not allow to discriminate if the autocorrelations in the turning angles are due to path integration, or more simply they reflect correlations in the pheromone map experienced by the ants.

Antennal contacts can also help an ant head toward the nest, or away from it, by sensing whether or not colony mates travelling on the same trail have recently come out of the nest. Another aspect that is not reproduced in our model but is often seen in ant trails is branching. Surprisingly, however, trail branching is not particularly prevalent in the arena-level observations. The branching that does occur is maybe explainable only in terms of trails ``colliding'' instead of an active formation of bifurcations. We can speculate that crowding and ant interactions on the trail are the main factors that determine branching~\cite{Dussutour_et_al2004}. Once formed, we would expect that Weber's Law response would aid the stability of branches. We can see this by noting that large $L + R$ in equation~\ref{eq:erf} results in each branch of a trail being nearly equally likely to be chosen even in the presence of small differences between $L$ and $R$. 

To answer these remaining questions more detailed arena-level observations of trail formation are needed, combined with detailed observations of how the ants interact with each other on the trails. We believe that the computer-automated approach we have used here can be further refined to produce such analyses.

\section*{Methods}
\subsection*{Experiments and Data Analysis}
We used colonies of the Argentine ant \textit{Linepithema humile} (Formicidae, Dolichoderinae) collected on the University Campus in Toulouse, France. In the south of France, Argentine ants are an invasive species that forms a single gigantic colony extending along the Mediterranean coast from Italy to Portugal~\cite{Vogel_et_al2010}. Ants were housed in artificial plaster nests and reared in an experimental room at a constant temperature of $26 \deg$ C under constant light conditions (L:D 12:12) and fed ad libitum with a mixture of eggs, carbohydrates, and vitamins~\cite{Bhatkar_and_Whitcomb1970} and with \textit{Musca domestica} maggots. Twelve groups containing 1,000 workers each, no queen and no brood were counted about one week prior to the experiments and placed in separate nests 10 cm in diameter, connected to a small foraging arena also 10 cm in diameter. These groups were starved for 24 hours before each experiment to stimulate exploratory behaviour.

Argentine ants leave pheromone both when moving out of the nest in search for food and when going back to the nest in food recruitment \cite{VanVorhisKey_and_Baker1986} as well as during exploration~\cite{Aron_et_al1989}. In the experiment, the ants were left free to explore an initially unmarked circular arena (diameter 1m). The arena was enclosed by 20 cm high walls covered with Fluon$^{\tiny \textregistered}$ to prevent ants from escaping. The floor of the arena was covered with a sheet of chlorine free paper, replaced after each trial. The entire setup was surrounded with white homogeneous curtains to eliminate as many orientation cues as possible. No food source was present in the arena at any time. Prior to the beginning of each trial, one colony was put under the arena and given access to it through a plastic tube opening on the arena centre. Each trial began when the first ant enters the arena and lasted for one hour. We conducted 12 such trials in total.

Under these conditions, Argentine ants start exploring the arena homogeneously around the entrance, but soon end up forming trails, some of which are then abandoned after some time and some of which are reinforced (figure~\ref{fig:snapshots}).

Two sets of data were collected for each trial. Snapshots of the whole arena were collected every 1s with a digital photo camera (Canon EOS 20D) and stored as 3504x2336 pixels RGB colour images. Image quality was sufficient to clearly see all the ants in the arena. These images were used for all arena-level observations of trail formation. At the same time, a smaller portion of the arena $\sim 47 \textrm{x} 38$ cm was filmed at 25~FPS and $720 \textrm{x} 576$ pixel resolution. The videos offered the additional temporal resolution necessary to analyse individual-level behaviour and were used for all quantifications of individual ant movement and response to pheromone. Each trial involved the recording of 3600 snapshots of the whole arena and 90000 video frames. 

The positions of all the ants were detected from each frame (and from each camera snapshot) with standard image analysis techniques (see e.g.~\cite{Parker1997, Gonzalez_and_Woods2002}: subtraction of a reference image, binarization (by thresholding independently the red, green and blue channel to detect pixels that were significantly darker than in the corresponding reference image for all channels), detection and labelling of connected components. Each connected component usually represented a single ant, but in rare cases it could mark two (and very rarely more) ants close together. We could easily detect the number of ants in a component by comparing its size with the typical size of single ants on the image (estimated as the mode of the number of pixels in each connected component detected throughout the experiment).

Pheromone levels cannot be measured directly in our experimental setup. Instead, as a proxy for the concentration of ``pheromone'' at each particular point in the arena we use the number of passages of ants over that point. Here we used only data from the individual-level film camera. An empty ``pheromone map'', corresponding to the field of view of the camcorder was initialized at the beginning of each trial. Then, at each frame, the ``pheromone'' of all the pixels that were covered by one ant was incremented by a fixed amount $\delta$, where $\delta$ is chosen in such a way that every ant marked on average the equivalent of one square millimetre of surface with one unit of pheromone every one second. This scaling with $\delta$ is purely arbitrary, but it allows us to compare between different trials. Supplementary movie 1 illustrates the outcome of the above process on a sample movie. The pheromone map $M$ at time $t$ on site $(x,y)$ is defined as
\begin{equation}
M \left( x,y,t \right) = \delta \sum\limits_{\tau = 0}^t 2^{- \frac{t - \tau}{\lambda}}C \left( x,y,\tau \right)
\label{eq:pheromone_map}
\end{equation}
where $C \left( x,y,t \right) = 1$ if patch $(x,y)$ is covered by (at least) one ant at time $t$ and 0 otherwise. We explore two possible scenarios: in the first scenario ($\lambda = \infty$) the pheromone does not undergo evaporation throughout the duration of the trial. The second scenario is one in which the pheromone evaporates with a half-life time $\lambda = 30 \textnormal{min}$. The choice of these two different scenarios is motivated by previous results by Van Vorhis Key and collaborators. In particular, from the data reported in~\cite{VanVorhisKey_and_Baker1982} it appears that the synthetic argentine ant pheromone (Z)-9-hexadecenal is active with a half-life of the order of 30 minutes. However, when gaster extracts obtained directly from ants were used~\cite{VanVorhisKey_et_al1981} the compound was active for a much longer time, with a half-life of the order of 4 hours. So, 30~min is a lower bound for the half-life time of pheromone evaporation. The results with and without pheromone evaporation are similar; when not specified our figures report results in the condition without evaporation.

In order to characterise the movement of ants we automatically tracked all the ants in the individual level videos over short time periods. This involves about 50,000 tracking events for each experimental trial, for a total of $\sim 600,000$ tracking events throughout the whole experiment. Every twenty frames (0.8 seconds) we mark the positions of all the ants in the field of view of the camera (see supplementary movie 2), follow them for ten frames, mark their positions again, follow them for ten more frames and mark their final positions. Occasionally, ants moved very little (less than 0.4 cm) in either of the two 10 frames intervals. Given the difficulty of estimating directions of movement and turning angles in these conditions, these data were discarded from all the analyses (except for computing the distribution of ant speeds in figure~\ref{fig:speed}A). This provides us with a simplified description of ant movement in terms of two straight segments and a turning angle. (See supplementary movie 2 for a visual representation of these operations).

We looked at the total pheromone within a one centimetre radius of the ant. In particular, we define L to be the total pheromone in a 90 degrees front-left sector relative to the ant’s position (see figure \ref{fig:regions_R_and_L} and supplementary movie 2). Similarly, $R$ is the total pheromone in a 90 degrees front-right sector. During an exploratory phase of data analysis we investigated multiple sectors with angles of 45 degrees and/or larger radius. The results were robust to such changes. 

\subsection*{Simulation}

In order to test the ability of the individual response rules to the pheromone concentration to explain trail formation at a larger scale, we set up an agent-based model of Argentine ant arena exploration using the NetLogo 4.1.1 modelling environment. The NetLogo world consists of square patches each with its own pheromone concentration. We run the simulation both in an open arena setup (diameter = 1000 patches) and in a ``binary bridge'' setup (total length of the setup 440 patches, all other proportions as in figure \ref{fig:results_double_bridge}A, with one nest at one extremity of the bridge and a ``food source'' at the opposite extremity. Parameters for patch size (one patch = 1 millimetre) and time step (one time step = 0.1 seconds) were chosen so as to provide a convenient scaling with the experiments.

1000 ants are initialized inside the nest at the beginning of the simulation; each ant enters the arena with a probability of 1/1000 per time step. Once in the arena, ants move at a constant speed (2 patches per time step, equal to the average speed of ants measured in the experiment). Their movement is not bound to the dimensions of the patches (ants move off-lattice), while pheromone concentrations are updated at the grain size of the grid, as pheromone is a property of the patches themselves. The ants move at every time step and update their direction of movement every four time steps (equivalent to 0.4s, the same interval used when analysing the data). The new direction of movement is determined by the concentration of pheromone within two circular sectors oriented 45 degrees to the left $(L)$ and to the right $(R)$ of the ant’s position. Each sector subtends an angle of 45 degrees, with a radius of 20 patches (2~cm). The angle of direction change $\alpha$ is then given by
\begin{equation}
\label{eq:simulation}
\alpha = \frac{\left( L-R+\epsilon_1 \right)}{\left(L+R\right)} + \epsilon_2
\end{equation}
where A is determined by fitting the data and $\epsilon_1$ and $\epsilon_2$ are random normal variables distributed with mean 0 and standard deviations $\sigma_1$ and $\sigma_2$, respectively. In our simulations $\sigma_1 = 50$ and $\sigma_2 = 15$. Once an ant moves to a new location, it leaves 0.1 units of pheromone at that location before making its next move. 

All the simulation parameters are chosen to match as closely as possible the experimental measures on real ant behaviour. We do not have experimental data to characterise the behaviour of ants along the arena border. In the simulation, ants heading against the border align with it, pointing to the direction that involves the minimum change from previous direction. In the binary bridge setup, ants heading against the border align with the border pointing away from the latest visited site (either nest or food source).

Ants that have been out in the arena for a long period can be marked with a special ``back-to-nest'' label. If such a labelled ant happens to be within two centimetres of the nest, then it will set its heading towards the nest and go directly there. In the open arena simulation, the ants are given a “back-to-nest” label randomly, with probability of 1 every 10 minutes of simulation time; in the binary bridge setup, the label is given to all the ants that have reached the food source at the distal extremity of the bridge (the dark blue region in figure~\ref{fig:results_double_bridge}A).

The simulations implement a very simplistic model of ant behaviour: ants respond only to local concentrations of pheromone, with no memory of past position and direction of movement. There are no direct ant-ant interactions. As such their purpose is to test to what degree the observed pheromone trail patterns are explainable simply in terms of their reaction to pheromone concentrations using equation~\ref{eq:simulation}.

\subsection*{Relationship between Weber's law and the Deneubourg model}

\label{sub:derivation_erf}

In the introduction, we stated that equation~\ref{eq:Weber} is similar to equation~\ref{eq:Deneubourg} with $a = 1$. To see this, impose $a=1$ (and $h=0$) in equation~\ref{eq:Deneubourg} we obtain
\begin{equation}
P_L = \dfrac{L}{L+R} \; \textnormal{and} \; P_R = \dfrac{R}{L+R}
\label{eq:P_L}
\end{equation}
Assume that a trail-following ant experiences pheromone concentration $L$ to its left and $R$ to its right and it moves to the left or to the right with probability $P_L$, respectively $P_R$. After following this procedure for $c$ time steps the expected position of the ant is equal to 
\begin{equation}
c\left( P_L - P_R\right) = c \dfrac{L-R}{L+R}
\label{eq:Michelson}
\end{equation}
which has the same form as equation~\ref{eq:Weber}.

In the derivation of equation~\ref{eq:erf} for the double bridge experiment the question is not about expected position, but rather the probability of ants arriving at either side of a branching point on the bridge. Consider ants which move up a single bridge toward a branching point. We can model the time evolution of the probability density $P(x,t)$ of an individual to be at a particular position $x$ as
\begin{equation}
\label{eq:drift_diffusion}
\frac{\partial P(x,t)}{\partial t} = - D_1 \frac{\partial P(x,t)}{\partial x} + D_2 \frac{{\partial}^2 P(x,t)}{\partial x}
\end{equation}
where $D_1$ and $D_2$ are respectively the drift and the diffusion coefficients. In our case the drift is, from equation \ref{eq:Michelson}, equal to
\begin{equation}
\label{eq:drift}
D_1 = \frac{R - L}{L + R}
\end{equation}
and we assume it to be constant for the whole duration of the walk in the decision zone between the left and right branches of the bridge. We further assume the diffusion, $D_2$, to be constant.

The initial condition is a delta function at zero, $\delta(x )$, so that initially the ant enters the middle of the bridge. The solution of equation~\ref{eq:drift_diffusion} subject to this initial condition is
\begin{equation}
P(x,t) = \frac{1}{\sqrt{4 \pi D_2 t}} \operatorname{exp} \left( - \dfrac{\left( x - D_1 t\right)^2}{4 D_2 t} \right)
\end{equation}
The survival probability, that is, the total probability at time $t$ of finding the individual in the region from $-\infty$ to $X$ is
\begin{equation}
\label{eq:survival_probability}
\begin{split}
S(X,t) = \int_{- \infty}^{X} P(x,t) dx 
= \frac{1}{\sqrt{4 \pi D_2 t}} \int_{- \infty}^{X} \operatorname{exp}\left( - \dfrac{\left( x - D_1 t\right)^2}{4 D_2 t} \right) dx
\end{split}
\end{equation}
We are interested in the probability that the ant ends up to the left of its starting point, i. e. $X=0$, which, after integrating, is given by
\begin{equation}
S(0,t) = \left[ 1 + \operatorname{erf} \left( \dfrac{ D_1 \sqrt{t}} {\sqrt{4 D_2}} \right) \right]
\end{equation}
where $ \operatorname{erf}$ is the error function. Since the time $t$ and variance $D_2$ are arbitrary parameters we can without loss of generalization set them both equal to unity and obtain
\begin{equation}
1 + \operatorname{erf} \left(  \frac{R - L}{L + R}  \right)\label{finalerf}
\end{equation}
Like equation \ref{eq:Deneubourg} with $a>1$, this equation is a sigmoidal function of $D_1$.

\subsection*{Bifurcation diagram for ants on a binary bridge}
\label{sub:bifurcation_diagram}
Let $y_1$ and $y_2$ be the densities of ants on each of the two branches $\textnormal{Y}_1$ and $\textnormal{Y}_2$ of a binary bridge setup (figure~\ref{fig:results_double_bridge}A). The evolution equation for the density of ants on each branch of the bridge can be written as 
\begin{equation}
\label{eq:double_bridge_density_evolution1}
\frac{d y_1}{dt} = \varphi f \left( y_1, y_2 \right) - y_1
\end{equation}
\begin{equation}
\label{eq:double_bridge_density_evolution2}
\frac{d y_2}{dt} = \varphi f \left( y_2, y_1 \right) - y_2
\end{equation}
The first positive terms in equations~\ref{eq:double_bridge_density_evolution1} and~\ref{eq:double_bridge_density_evolution2} correspond to the traffic on $\textnormal{Y}_1$ and $\textnormal{Y}_2$ respectively. Here $\varphi$ is the total flux of individuals per unit time and $f$ is the probability of taking a particular branch. The negative term corresponds to the spontaneous retirement of ants from each branch. The particular form of $f$ chosen here is consistent with equation \ref{finalerf}, i.e.
\begin{equation}
\begin{split}
f \left( y_1, y_2 \right) = \frac{1}{2} \left[ 1 + \operatorname{erf} \left( \dfrac{y_1 - y_2}{y_1 + y_2 + T_0}\right) \right] \\
f \left( y_2, y_1 \right) = \frac{1}{2} \left[ 1 + \operatorname{erf} \left( \dfrac{y_2 - y_1}{y_1 + y_2 + T_0}\right) \right]
\end{split}
\end{equation}
The additional parameter $T_0$ captures the threshold level of response found in the experiment (see figure \ref{fig:slope_vs_L_plus_R}).

At the stationary state, putting $d y_1 / dt = 0$, $d y_2 / dt = 0$ and adding equation \ref{eq:double_bridge_density_evolution1} and \ref{eq:double_bridge_density_evolution2} we have 
\begin{equation}
\label{eq:phi_stationary_state}
\varphi = y_1 + y_2
\end{equation}
replacing~\ref{eq:phi_stationary_state} in~\ref{eq:double_bridge_density_evolution1}
\begin{equation}
\frac{\varphi}{2} \left[ 1 + \operatorname{erf} \left( \dfrac{2 y_1 - \varphi}{\varphi + K} \right)\right] - y_1 = 0
\end{equation}
which can be solved numerically.

Figure~\ref{fig:results_double_bridge}D gives the bifurcation diagram of $y1 / (y_1 + y_2)$ against $\varphi$. We see that the figure displays a pitchfork bifurcation, reminiscent of the behaviour of Deneubourg's model on a bridge experiment with two equal paths: the homogeneous state, stable for small values of flow, loses its stability at a critical value where two new inhomogeneous solutions appear. The stability has been checked by integrating equations \ref{eq:double_bridge_density_evolution1} and \ref{eq:double_bridge_density_evolution2} numerically.

\section*{Acknowledgments}
This work was supported by a European Research Council starting grant to D.J.T.S. (ref: IDCAB). The experimental work was performed in Toulouse under funding from the French ANR Project MESOMORPH (grant ANR-06-BYOS-0008). Simon Garnier is grateful to Iain Couzin for fruitful and inspiring discussion.

\bibliography{Linepithema1.bib}

\section*{Tables}
\begin{table}[!ht]
\caption{
\bf{Fit of eq.~\ref{eq:power_law} parameters from individual replicates; pheromone evaporation is not assumed}}
\begin{tabular}{|c|c|c|c|c|c|}
Trial I.D. & $A (95\% C.I.)$ & $\beta (95\% C.I.)$ & $R^2$ & $F$ & $P val.$ \\
T01 & $30.94 (19.66, 48. 68)$ & $1.020 (0.946, 1.094)$ & $0.9816$ & $F_{16,1} = 856$ & $p < 0.001$ \\
T02 & $39.65 (26.32, 59.73)$ & $1.034 (0.968, 1.101)$ & $0.9854$ & $F_{16,1} = 1077$ & $p < 0.001$ \\
T03 & $33.99 (27.85, 41.48)$ & $1.031 (0.999, 1.064)$ & $0.9965$ & $F_{16,1} = 4530$ & $p < 0.001$ \\
T04 & $30.62 (21.36, 43.90)$ & $1.014 (0.954, 1.074)$ & $0.9886$ & $F_{15,1} = 1299$ & $p < 0.001$ \\
T05 & $34.38 (26.05, 45.37)$ & $1.031 (0.983, 1.079)$ & $0.9940$ & $F_{13,1} = 2136$ & $p < 0.001$ \\
T06 & $56.84 (41.54, 77.78)$ & $1.097 (1.043, 1.152)$ &  $0.9932$ & $F_{13,1} = 1893$ & $p < 0.001$ \\
T07 & $48.90 (39.04, 61.26)$ & $1.085 (1.048, 1.122)$ & $0.9959$ & $F_{16,1} = 3922$ & $p < 0.001$ \\
T08 & $34.52 (20.37, 58.47)$ & $1.006 (0.914, 1.097)$ & $0.9774$ & $F_{13,1} = 563$ & $p < 0.001$ \\
T09 & $40.86 (30.45, 54.81)$ & $1.032 (0.982, 1.082)$ & $0.9929$ & $F_{14,1} = 1960$ & $p < 0.001$ \\
T10 & $53.62 (28.40, 101.25)$ & $1.090 (0.972, 1.208)$ & $0.9769$ & $F_{10,1} = 423$ & $p < 0.001$ \\
T11 & $23.38 (8.85, 61.76)$ & $0.990 (0.821, 1.159)$ & $0.9250$ & $F_{13,1} = 160$ & $p < 0.001$ \\
T12 & $42.96 (20.00, 92.28)$ & $1.023 (0.885, 1.162)$ & $0.9598$ & $F_{11,1} = 263$ & $p < 0.001$ \\
\end{tabular}
\begin{flushleft}The table reports for each trial the values of a power law fit of the type $k = A\left(L+R\right)^{- \beta}$, where $k$ is the slope of the angle change vs. pheromone difference $k = \frac{\alpha}{(L-R)}$. Fit values are obtained through non-weighted linear least squares fit of the log-transformed data.
\end{flushleft}
\label{tab:no_evaporation}
\end{table}

\begin{table}[!ht]
\caption{
\bf{Fit of eq.~\ref{eq:power_law} parameters from individual replicates when assuming pheromone evaporation}}
\begin{tabular}{|c|c|c|c|c|c|}
Trial I.D. & $A (95\% C.I.)$ & $\beta (95\% C.I.)$ & $R^2$ & $F$ & $P val.$ \\
T01 & $31.91 (21.36, 47.69)$ & $1.030 (0.960, 1.101)$ & $0.9835$ & $F_{16,1} = 954$ & $p < 0.001$ \\
T02 & $35.13 (23.31, 52.96)$ & $1.020 (0.951, 1.089)$ & $0.9816$ & $F_{16,1} = 961$ & $p < 0.001$ \\
T03 & $26.30 (19.44, 35.58)$ & $0.991 (0.938, 1.045)$ & $0.9898$ & $F_{16,1} = 1559$ & $p < 0.001$ \\
T04 & $30.07 (22.84, 39.59)$ & $1.009 (0.958, 1.059)$ & $0.9924$ & $F_{1,41} = 1821$ & $p < 0.001$ \\
T05 & $42.48 (35.46, 50.87)$ & $1.072 (1.038, 1.106)$ & $0.9972$ & $F_{13,1} = 4627$ & $p < 0.001$ \\
T06 & $55.36 (36.89, 83.08)$ & $1.104 (1.024, 1.184)$ &  $0.9881$ & $F_{11,1} = 913$ & $p < 0.001$ \\
T07 & $58.67 (34.19, 100.68)$ & $1.138 (1.043, 1.234)$ & $0.9758$ & $F_{16,1} = 644$ & $p < 0.001$ \\
T08 & $38.03 (25.59, 56.51)$ & $1.026 (0.952, 1.101)$ & $0.9854$ & $F_{13,1} = 867$ & $p < 0.001$ \\
T09 & $32.82 (22.31, 48.28)$ & $0.993 (0.920, 1.066)$ & $0.9852$ & $F_{13,1} = 867$ & $p < 0.001$ \\
T10 & $46.96 (29.62, 74.45)$ & $1.078 (0.985, 1.172)$ & $0.9851$ & $F_{10,1} = 659$ & $p < 0.001$ \\
T11 & $26.16 (12.28, 55.70)$ & $1.007 (0.865, 1.150)$ & $0.9471$ & $F_{13,1} = 233$ & $p < 0.001$ \\
T12 & $33.99 (11.68, 98.92)$ & $0.990 (0.773, 1.207)$ & $0.9118$ & $F_{10,1} = 103$ & $p < 0.001$ \\
\end{tabular}
\begin{flushleft}The same as table 1, but assuming pheromone evaporation with a half-life of 30 minutes ( $\lambda = 30$ in equation \ref{eq:pheromone_map}).
\end{flushleft}
\label{tab:with_evaporation}
\end{table}

\clearpage
\newpage

\section*{Figure Legends}

\renewcommand{\figurename}{Sup. Video} 
\setcounter{figure}{0} 

\begin{figure}[!ht]
\begin{center}
\end{center}
\caption{
{\bf Video illustrating how the pheromone maps are inferred from the movement of the ants. } Whenever a portion of the arena is covered by one ant for one frame, the pheromone map at that location is incremented by a constant amount. Notice that the movie is made for illustrative purposes: real experiments involved filming a larger portion of the arena from above and from the very beginning of the trial; pheromone quantities were stored with 24 bit precision.
}
\label{supplementary_video_1}
\end{figure}

\begin{figure}[!ht]
\begin{center}
\end{center}
\caption{
{\bf Video illustrating how the movement of the ants is analysed and correlated to the pheromone maps.}  Every twenty frames (0.4 s) each ant present under the field of view of the camcorder is detected and its position marked. The ant is followed over two intervals of 10 frames each and its path is described by two paths of ten frames each and a turning angle $\alpha$. These measures of ant movement are stored together with the whole pheromone map around the ant position in correspondence with the turning angle (more precisely with the map computed 16 seconds before the tracking event). This analysis is repeated every twenty frames for the whole duration of the arena level movies. Only the tracking events for which the ant was clearly detected in all the frames and did not meet any other ant were kept for the analyses. 
}
\label{supplementary_video_2}
\end{figure}

\end{document}